\documentclass[12pt,preprint]{aastex}
\slugcomment{Submitted to: 
The Astrophysical Journal}
\shorttitle{Transparent Sun}
\shortauthors{Nemiroff, Patla}

\begin{document}

\title{Gravitational Lensing Characteristics of the Transparent Sun}

\author{ Bijunath Patla and Robert J. Nemiroff }
\affil{Michigan Technological University, Department of Physics, \\
1400 Townsend Drive, Houghton, MI  49931}

\begin{abstract} 
The transparent Sun is modeled as a spherically symmetric and centrally condensed gravitational lens using recent Standard Solar Model (SSM) data.  The Sun's minimum focal length is computed to a refined accuracy of 23.5 $\pm 0.1$ AU, just beyond the orbit of Uranus.   The Sun creates a single image of a distant point source visible to observers inside this minimum focal length and to observers sufficiently removed from the line connecting the source through the Sun's center.  Regions of space are mapped where three images of a distant point source are created, along with their associated magnifications.  Solar caustics, critical curves, and Einstein rings are computed and discussed.  Extremely high gravitational lens magnifications exist for observers situated so that an angularly small, unlensed source appears near a three-image caustic.  Types of radiations that might undergo significant solar lens magnifications as they can traverse the core of the Sun, including neutrinos and gravitational radiation, are discussed.
\end{abstract}


\section{Introduction}
Our Sun is known to act as a gravitational lens.  The angular shift of the apparent position of a star located behind the Sun was first observed during the Solar eclipse of 1919 \citep{eddington}, in conformity with the predictions of general relativity \citep{einstein16}.  Years later, the lens like action of distant stars was discussed more generally  \citep{chwolson,einstein36}.  The influence of that discussion has caused the ring seen by the observer during perfect alignment of the source, lens and the observer to be called the Einstein ring around a point lens. For a detailed historical review, see for example, Wambsganss (1998).

Parallel light rays incident with an impact parameter $b$ on the lens plane gets deflected by an angle $\alpha$. The deflection angle is inversely proportional to the impact parameter $b$ according to the formula first derived by Einstein (1916)
\begin{equation}
\alpha=\frac{4G M(b)}{R_\odot b c^2}  ,
\label{def}
\end{equation}
where $b$ is the dimensionless impact parameter of the passing light ray, $M(b)$ is the deflecting mass, $G$ is the gravitational constant and $c$ is the speed of light in vacuum.  $M_\odot$ and $R_\odot$ are the mass and radius of the Sun respectively. Assuming the deflection angle is small, the minimum focal length for the opaque Sun is approximated by 
\begin{equation}
F=\frac{R_\odot^2 c^2}{4 G M_\odot}=548.30(\pm 0.01)\;\rm{AU} ,
\label{focal}
\end{equation}
where F is the minimum focal length of the opaque Sun.
The uncertainty in F is based on statistical errors on $R_{\odot}$ and $G M_{\odot}$ \citep{cox}.  Slight offsets of the lens or source from the optic axis will break the Einstein ring into two bright images as seen by the observer. 

The possibility of using the  transparent Sun as a gravitational lens was previously discussed by various authors \citep{lawrence, clark, ohanian, cyranski, bontz, burke85, nemiroff97, demkov, escribano}. Burke (1985) computed the minimum focal length to be 25 AU, by considering the maximum value of the deflection angle, as a function of the radius of a cylindrical mass. Demkov and Puchkov (2000), on the other hand, computed a minimum focal length of the transparent Sun to be about 24 AU.  Starting with the Standard Solar Model as it was known in 1989 \citep{bahcall89}, they computed the gravitational lens deflection angle as a function of impact parameter and then approximated this with a Taylor's series near the Sun's center.   

Throughout the present analysis, the more recent Bahcall et al.(2005) Standard Solar Model (SSM) data is used.   A more complete model of the gravitational lens characteristics of the transparent Sun is computed, including a more accurate minimum focal length (23.5 $\pm$ 0.1 AU), regions of multiple images, and the locations of caustics and critical curves.  

The plan of the paper is as follows: $\S$ 2 explains the connection between the power-law density profiles of the stars to the deflection angles produced due to their lens action. $\S $ 3 outlines two alternative approaches to obtain the minimal focal length of the transparent Sun. The strength of a composite lens capable of producing multiple images is given a through treatment in $\S$ 4. The critical curves, caustics, magnification and multiple image zones are computed in $\S$ 5, and conclusions summarized in $\S$ 6.

\section{Minimum focal length of transparent stellar lenses}
Transparent stellar lenses were previously studied in the context of gravitational radiation \citep{lawrence,ohanian}. The lens action of a simple model of the transparent Sun to gravitational radiation was studied by Bontz and Haugan (1973), and a uniform transparent lens was analyzed by Clark (1973).  One characteristic of a transparent lens is its minimum focal length, defined as the {\it minimum} distance between the center of the lens and the point on the optic axis where the deflected rays corresponding to different impact parameters converge.

Main sequence stars, like our Sun, are in a state of hydrostatic equilibrium.  When describable by a single hydrostatic state, their pressure and density are simply proportional to their radius raised to a polytropic (power-law) index \citep{chandra}.  It can be shown generally that gravitational lens deflection angles fall off as $r^{n+1}$ when the density falls as polytropically as $r^{n}$ \citep{burke85}.  Outside of a certain radius, the density profile of Sun-like stars is well characterized by a single polytropic index such that it falls off nearly as $1/r^2$.

To elucidate the general problem of the solar focal length, an instructive exercise could be to compare various idealized power-laws with the Standard Solar Model data.
Consider a star of mass $M_*$ and radius $R_*$. Let $\rho(r)$ be the idealized density profile of the star. 
\begin{equation}
\rho(r)=\rho_o r^n ,
\label{den1}
\end{equation}
where $\rho_o$ is a constant, r is the radial distance from the center of the star and n is the power-law index. The value of $\rho_o$  is determined by the mass and radius of the star. The normalized projected mass as a function of dimensionless impact parameter b, for power-laws corresponding to n = -2, -1 and 0 can be computed analytically by integrating equation (\ref{den1}) using a volume element of a cylindrical coordinate system. 
\begin{equation}
M(b)_{n=-2}/M_* = 1-\sqrt{1-b^2}+ b \tan^{-1}\sqrt{\frac{1}{b^2}-1} ,
\label{minus2}
\end{equation}

\begin{equation}
M(b)_{n=-1}/M_* =1-\sqrt{1-b^2}+b^2 \sinh^{-1}\sqrt{\frac{1}{b^2}-1},
\label{minus1}
\end{equation}

\begin{equation}
M(b)_{n=0}/M_* = 1-(1-b^2)^{3/2},
\label{zero}
\end{equation}
where M(b) is the mass enclosed within a cylinder of radius b. The corresponding expressions for the deflection angle and focal length are:
\begin{equation}
\alpha (b)_n=\frac{4 G M(b)_n}{c^2 R_*  b} ,
\label{reffocal}
\end{equation}
and

\begin{equation}
D(b)_n=\frac{R_* b}{\alpha (b)_n}=\frac{{R_*}^2b^2 c^2}{4 G M(b)_n},
\end{equation}
where $D(b)_n$ is the focal length and $\alpha(b)_n$ is the deflection angle corresponding to a power-law index n and impact parameter b. The function $M(b)_n$  is given by  equations (\ref{minus2}), (\ref{minus1}) and (\ref{zero}) for three different values for n.

\clearpage

\begin{figure}[h]
\begin{center}
\includegraphics[width=.70\textwidth]
{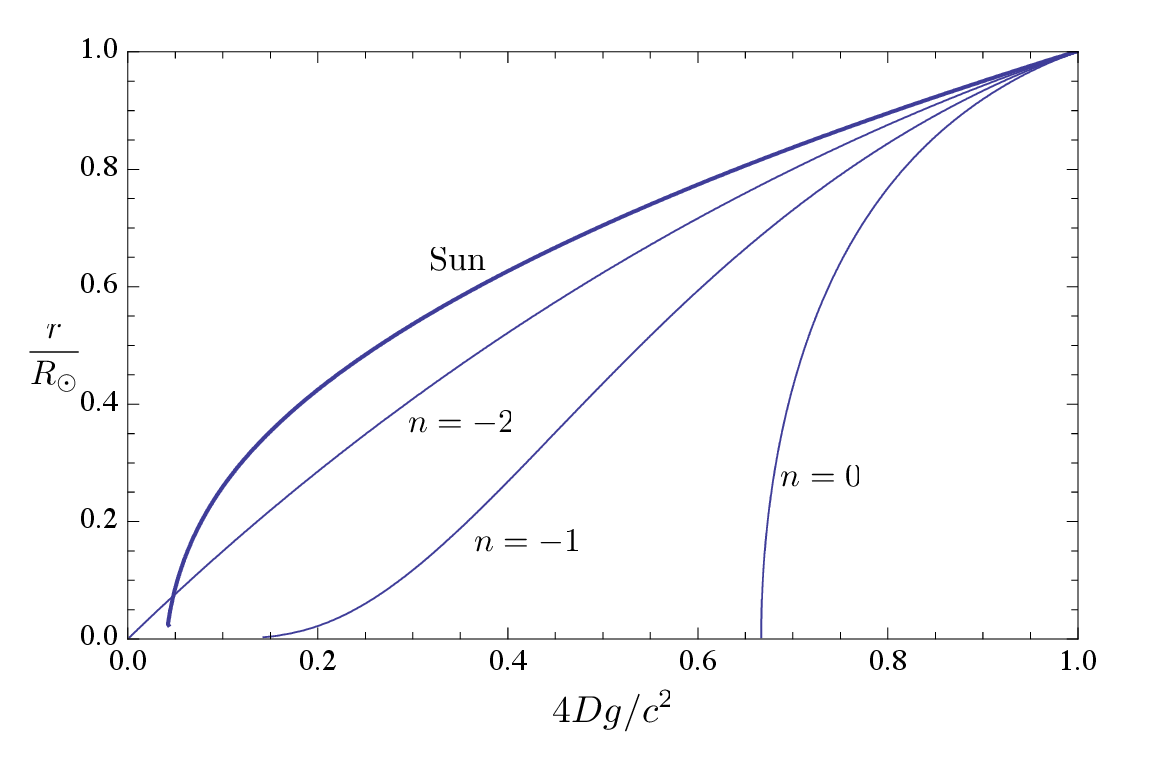}
\caption{The normalized focal length is plotted along the horizontal axis. The dimensionless impact parameter is plotted along the vertical axis. The n=0 curve corresponds to a constant density sphere with a minimum focal length of 365 AU. The Standard Solar Model data for the Sun sets  the minimum focal length of the Sun at 23.5 AU. The focal length for the Sun is computed numerically using the SSM data. }
\label{f1}
\end{center}
\end{figure}

\clearpage

A convenient relationship connecting the minimum focal length to an arbitrary index $n$ can be established for very small impact parameters by taking the limit b$\rightarrow$ 0 of equation (\ref{reffocal}),  using L' Hospital's rule. Introducing a constant  $g= G M_\odot /R_{\odot}^2$, using the mass and radius of the Sun,  a general expression for the limiting value of the focal length corresponding to  $n > -2$ may be obtained. Note that, $4 g /c^2$  corresponds to a length of $\approx$ 548 AU, the minimum focal length of the opaque Sun.
\begin{equation}
\lim_{b\rightarrow0} D(b)_n=D_n= \frac{c^2}{g}\frac{(1+n)}{2(3+n)},  
\label{referee}
\end{equation}
where $\lim D(b)_{b\rightarrow0}$ is {\it also the minimum focal length} for only n= -1 and n=0. For integer values of $n>1$, $\lim D(b)_{b\rightarrow0}$ is {\it no longer} the minimum focal length (see Fig.\ref{f1}). The homogenous sphere, with $n=0$, yields:
\begin{equation}
D_{n=0} = c^2/6g=\frac{2}{3}F,
\label{referee2}
\end{equation}
where $D_{n=0}$ and  F are the minimum focal length of a transparent uniform sphere and an opaque sphere of the same mass and radius respectively.  Equation (\ref{referee}) sets the minimum focal length of a transparent uniform sphere, with radius and mass equal to the radius and mass of the Sun, to be 365 AU \citep{lawrence,clark,ohanian}.  At $n =-1$, equation (\ref{referee}) goes to zero and for $n <-1$, it becomes invalid.

\begin{figure}[h]
\begin{center}
\includegraphics[width=.85\textwidth]
{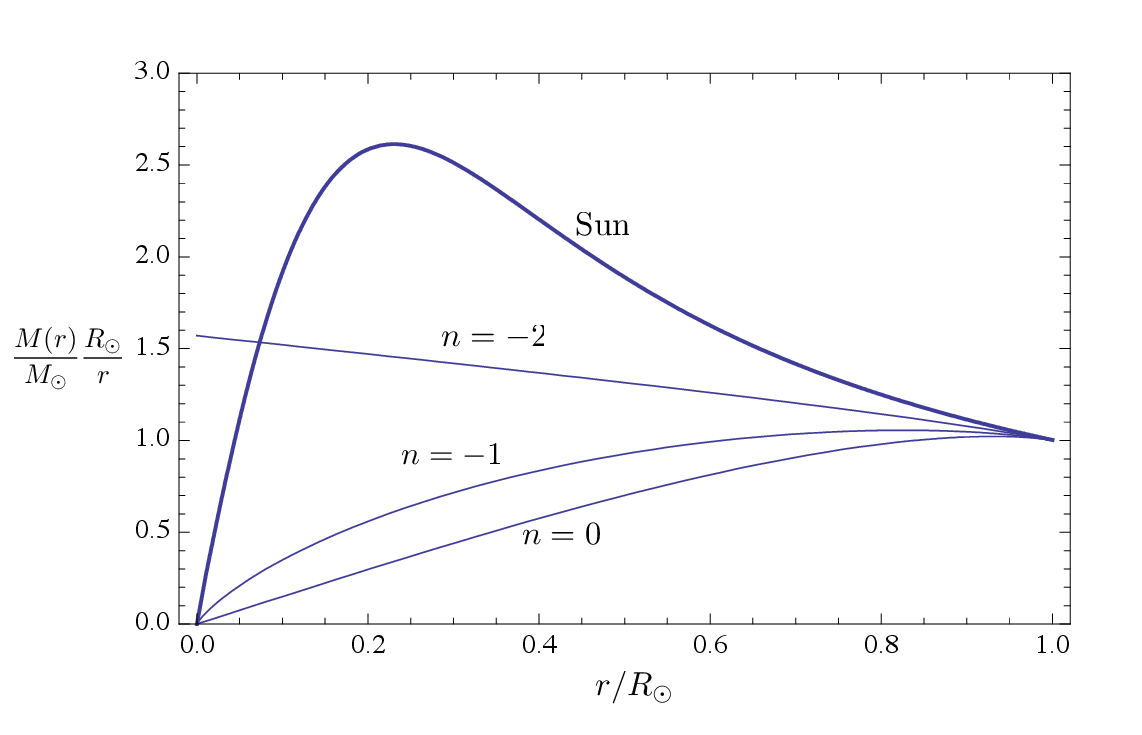}
\caption{The normalized M(r)/r, which is a measure of the deflection angle is plotted as a function of dimensionless impact parameter. The M(r)/r profile for the Sun rises toward the idealized n =-2 profile for small impact parameters before it starts falling off asymptotically.}
\label{f2}
\end{center}
\end{figure}

\clearpage

We note that the Sun roughly follows the n= -2 density profile outside of $\approx 0.075 R_\odot$. However, from Figure \ref{f1}, the focal length profile of the Sun mimics that of a constant density sphere within this impact radius, enclosing a mass of $\approx 0.10 M_\odot$. So, the value of the minimum focal length of the Sun can be crudely approximated to be 
\begin{equation}
F \approx \frac{c^2}{4g}\frac{(0.075)^2}{0.1}= 31 \rm{AU}.
\end{equation}

In contrast, in the following section, a more complex ray tracing method will be employed to obtain a more accurate estimate of the minimum focal length of the Sun.

\section{Solutions of the lens equation}

It is well known that only a point lens can generate two images in the weak deflection limit \citep{schneider}.  However, a composite lens can produce multiple images.  The location of images in the lens plane depends on the enclosed projected mass within the impact radius. Burke's odd number theorem \citep{burke} states that a composite lens can produce only odd number of images of the source, and the images always appear (disappear) in pairs as the source moves inside (outside) of a caustic. 

In the standard geometrical optics approximation, light rays start out straight from a source and then change directions discretely at the deflector plane, subsequently reaching the observer.  To utilize geometric optics for a composite lens, one must project the three-dimensional mass of the deflector onto a plane perpendicular to the optical axis connecting the source to the observer.  A plane wave approximation is assumed by considering source radiation that has a wavelength $\lambda \ll R_B$, where $ R_B$, is the  radius of curvature of the background spacetime.  

It is also assumed that the sources are point like and there is no contribution due to internally converging Ricci focusing by the mass within a pencil of rays. In other words the light bundle is assumed to be infinitesimally thin and the deflection is purely due to Weyl focusing by the projected mass lying within the impact radius \citep{dyer}. The mass outside of the impact radius does not contribute to the deflection, following Newton's and Birkhoff's theorem.

\clearpage

\begin{figure}[h]
\begin{center}
\includegraphics[width=.70\textwidth]
{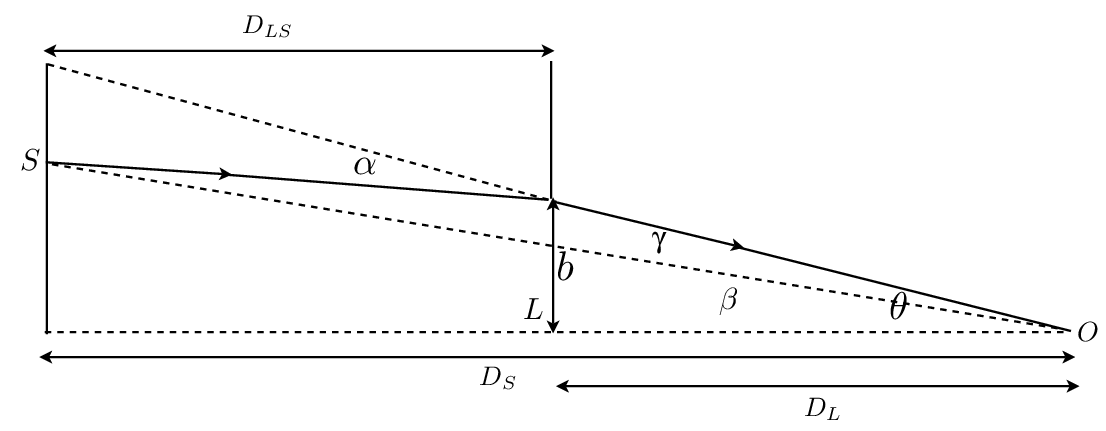}
\caption{The center of the lens $L$ is at a distance $D_L$ from the observer  $O$ and a distance $D_{LS}$ from the source $S$. $\beta$ and $\theta$ are the angles subtended by the unlensed source and its image visible to the observer.}
\label{schematic}
\end{center}
\end{figure}

\clearpage

The geometry of Figure \ref{schematic} satisfies the equation
\begin{equation}
\vec\beta=\vec\theta-\vec\alpha,
\label{lens}
\end{equation}
if one assumes $D_{LS}\approx D_S$, for distant sources.
$\beta$ and $\theta$ are the angles subtended by the unlensed source and its image visible to the observer. The deflection angle $\alpha$, same as equation (\ref{def}), can be recast in the form
\begin{equation}
\vec \alpha=\frac{4 G}{c^2}\int_{S} r\,d\phi dr\, \Sigma(r) \frac{\vec r}{|r|^2},
\label{lensden}
\end{equation}
where $\Sigma (r)$ is the value of the projected mass density at the point $r$ from the center of the lens, $r$ is the impact parameter, and $S$ the surface area in a polar ($r,\phi$) coordinate system . The center of the lens is made to coincide with the center of the preferred coordinate system for simplicity and owing to the cylindrical symmetry. 

The numerical solution of equation (\ref{lens}) yields the image locations. $\vec \alpha$ is a function of $\vec r$ (radius) only, owing to the circular symmetry. In general it is a function of two parameters of a polar coordinate system. In other words, the source and the image locations do not align on a straight line.
\clearpage

\begin{figure}[h]
\begin{center}
\includegraphics[width=.70\textwidth]
{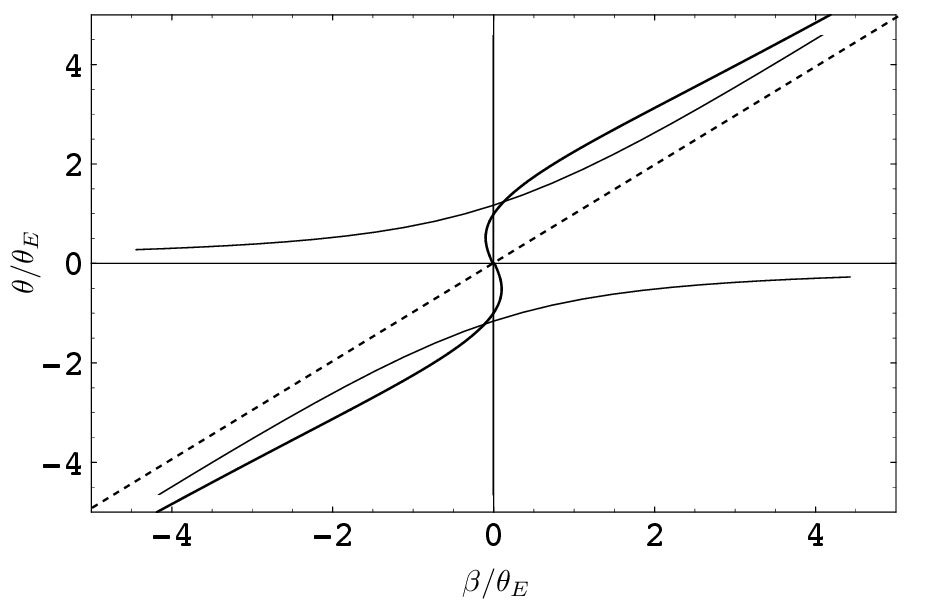}
\caption{The curves relate unlensed source positions $\beta$ and the image locations $\theta$ for three different scenarios. The continuous curve represents the solutions of the lens equation for a centrally condensed lens. The disjoint set of curves delineates a point lens of the same mass and the straight line corresponds to the absence of a lens, where unlensed source positions and image positions coincide at all times. The observer distance from the center of the lens is 50 AU.}
\label{betatheta}
\end{center}
\end{figure}

\clearpage

The solutions to the lens equation have a maximum of three roots. Any vertical line in the $\beta-\theta$ plane corresponds to a fixed source location, the points of intersection of which are the corresponding image locations.  Inspection of Figure \ref{betatheta} shows that sources far from the optic axis are seen as a single image.  The $\beta-\theta$ curves are conformal transformations for the mapping $\vec \theta \rightarrow \vec \beta$ for every observer location $D_L$. Note that Figure \ref{betatheta} is based on an inverted map that is used to calculate the magnification.

At $\beta=0$ the two images are equally separated from the optic axis. The assumed spherical symmetry of the lens will generate a circular Einstein ring on the lens plane. The formation of a real Einstein ring requires that a source be placed exactly along the line joining the observer through the center of the lens, which, for realistic cases, only occurs for sources of finite size.  Nevertheless, an Einstein ring still has theoretical significance since it separates sets of images \citep{nemiroff93}.  For example, as the source moves behind the lens, no source image will ever be seen to cross an Einstein ring. 
\clearpage

\begin{figure}[h]
\begin{center}
\includegraphics[width=.70\textwidth]
{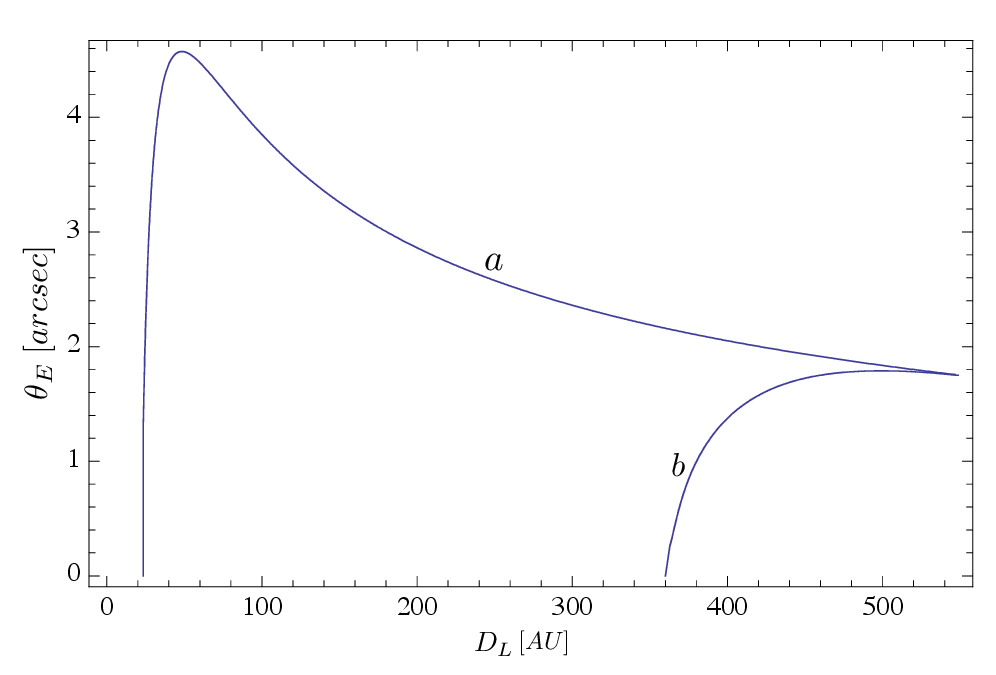}
\caption{The radius of Einstein rings as function of observer--lens distance $D_L$ along the optic axis. The curve ($a$) corresponds to the Sun and curve ($b$) corresponds to a homogenous lens with a radius and mean density equal to the Sun. The angle subtended at the limb of the Sun in both cases converge to $\approx 1.75"$}
\label{einstein}
\end{center}
\end{figure}

\clearpage

It is important to make a distinction between what is meant by focal length for an optical lens as opposed to a gravitational lens.  The focal length of an optical lens is the distance between the center of the lens and the point at which paraxial light rays converge or appear to diverge. The definition of deflection angle is 
\begin{equation}
b=D_L \tan\alpha,
\end{equation}
where $b$, $D_L$ and $\alpha$ are the impact parameter, focal length and deflection angle respectively. 

The gravitational lens has a focal length for each impact parameter bounded by a minimum focal length.  This is because the deflection angle is not a linear function of impact parameter. Paraxial light rays converge at different points along the optic axis.  

The existence of multiple images is a necessary condition for a consistent definition of focal length. Past the minimum focal length, the observed merger of two source images results in the formation of an Einstein ring.  Numerical simulations (discussed more in the Appendix) involving different values of $D_L$ and Burke's odd number theorem will impose the following mathematical condition, for the existence of multiple images  \citep{schneider}:
\begin{equation}
\frac{d\theta}{d\beta}<0.
\label{multiplecond}
\end{equation}
The least value of $D_L$ that conforms to the above condition, the minimum focal length, is found numerically to be $23.5$ AU. This implies that, for values of $D_L$ less than $23.5$ AU there are no multiple images and hence no focal point. The perfect alignment of the source, lens and the observer at a distance less than $23.5$ AU will not result in the formation of Einstein ring.

\section{Strength of a centrally peaked lens}

Not only does the Sun's mass density decrease monotonically with radius, but its mass density projected onto a lens plane also decreases monotonically with radius.  Surface density normalized to the critical surface density needed to generate multiple images is referred to as a dimensionless surface density (also called convergence) $\kappa$ and is a measure of the strength of the lens \citep{cowling,schneider}. 
\begin{equation}
\kappa (r)=\frac{\Sigma (r)}{\Sigma_{cr}},
\label{kappa}
\end{equation}
where
\begin{equation}
\Sigma_{cr}=\frac{c^2}{4\pi G D_L },
\label{critden}
\end{equation}
is the critical density and $D_L$ is the distance between the lens and the observer.
If $\kappa (\vec r )\ge 1$ for {\it some} regions on the lens then the lens is termed ``strong."  For instructive comparisons to our Sun, a ``constant density" sphere of radius $R_\odot$ with a volume density equal to the average density of the Sun is used.

A ``weak" lens is characterized by $\kappa (\vec r)\ll 1$ and cannot produce multiple images \citep{cowling}. The constant density sphere does not produce multiple images at $D_L=23.5$ AU. This is illustrated in Figure 4, where curve ($b$) has a value of $\kappa (\vec r)\ll 1$ throughout the entire the 
lens at 50.0 AU.  Weak lenses are only weak for observers sufficiently close -- observers farther than some minimal focal length will see the same object as a strong lens that creates multiple images.  In case of the constant density sphere of Solar mass, this value for $D_L$ is $\sim 365$ AU.

The validity of equation (\ref{multiplecond}) implies that the transition of the lens from being weak to strong should be traced by at least enough points that $\beta -\theta$ curve is continuous at the minimum focal length. Therefore, a ray tracing algorithm is employed to establish and verify the existence of a minimum focal length more precisely, with an error less than one percent, which is the combined error of the data \citep{bahcall05} and simulations.

\clearpage

\begin{figure}[h]
\begin{center}
\includegraphics[width=.70\textwidth]
{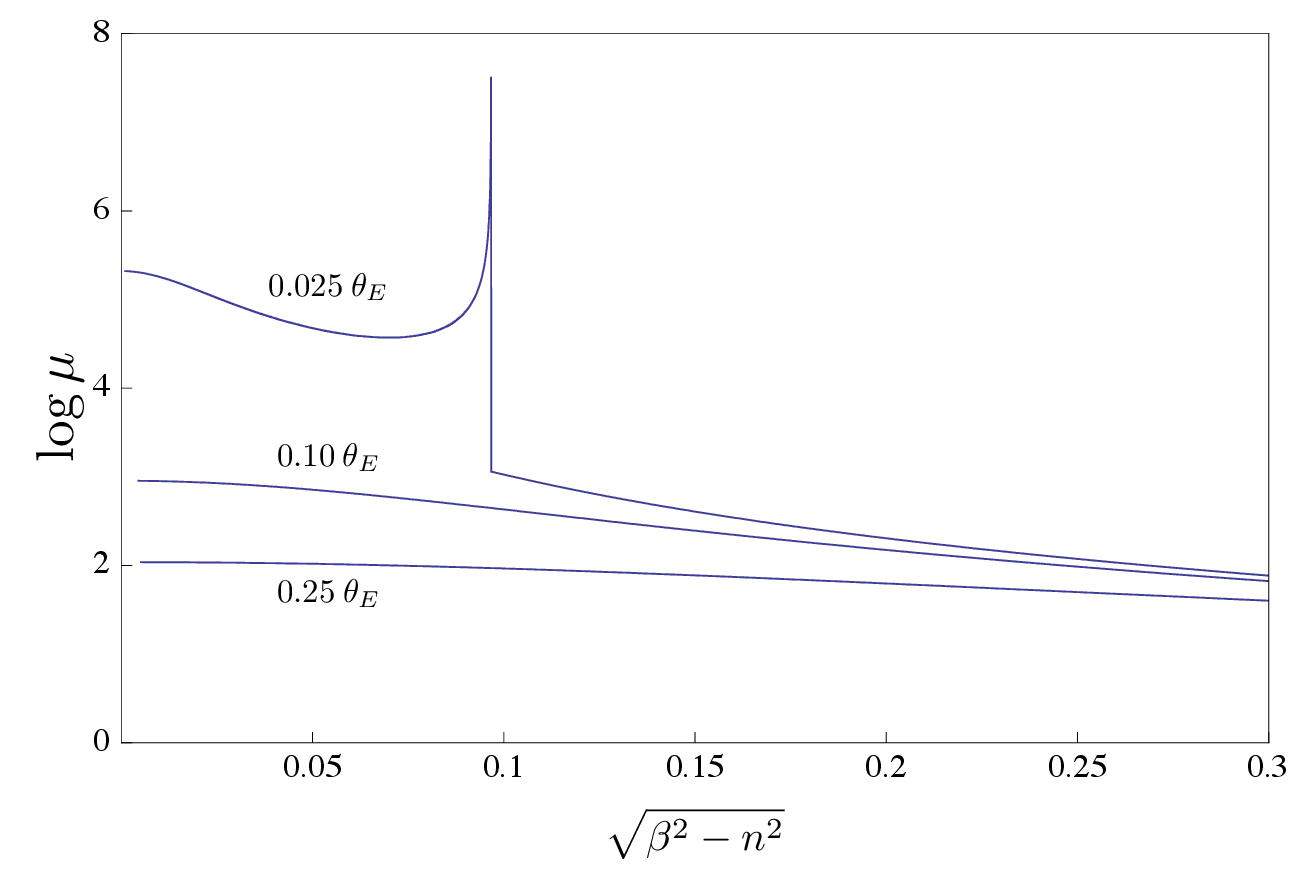}
\caption{The surface density $\kappa$ plotted on the vertical axis is a monotonically decreasing function of radius. 
The curves (a) corresponds to the Sun for observer--lens distance $D_L = 50$ AU, whereas curves (b) and (c) represents the homogeneous density sphere at $D_L$ of $50$ and $450$ AU.
The dimensionless surface density parameter suggests that the transparent Sun is a strong lens at $D_L = 50$ AU and a homogeneous sphere is a weak lens $\kappa \ll 1$ at the same $D_L$. But, the very same homogeneous sphere is capable of producing multiple images at observer--lens distance ($D_L$) greater than $365$ AU.  }
\label{figcritden}
\end{center}
\end{figure}

\clearpage

Alternatively, the minimum focal length of the Sun can be obtained by using the values of the mass enclosed and the impact parameter corresponding to the value of $\kappa (\vec r)=1$. Using equation (\ref{focal}) and the substituting the value $0.0135 M_\odot$ and the radius of the enclosed mass $.024 R_\odot$ one obtains a focal length of $\approx 23.5$ AU, which lies between the orbits of Uranus and Neptune. 

\section{Critical curves, caustics and magnification}

The lens plane is defined as the plane perpendicular to the line joining a point source and a fixed observer.  According to the odd number theorem for transparent gravitational lenses \citep{burke}, the observer will see an odd number of images of the point source no matter what position the center of the lens occupies in the lens plane.  Areas might exist in the lens plane where the lens center can be placed to create a specific odd number of images visible to the observer.  The boundaries between these areas are called critical curves. For a spherically symmetric lens, a critical curve is a circle. Were a lens center to lie on a critical curve, a formally infinite magnification of a point source would be seen by the observer.

The source plane is defined as a plane perpendicular to the line extending from a fixed observer through the lens center.  In analogy with the lens plane, areas might exist in the source plane where a point source can be placed so the (fixed) lens creates a specific odd number of images visible to the (fixed) observer.  The boundaries between these areas of the source plane are called caustics.  
For spherical lenses, the caustics are also circles.  Were a point source to lie on a caustic, a formally infinite magnification would be seen by the observer. For a given lens, the $\theta$ versus $\beta$ curve may show points of diverging slope.  These points correspond to infinite magnification, and hence yield the angular radii of the corresponding caustic ($\beta$) and critical ($\theta$) circles.  

Let's now consider specifically our Sun.  The $\theta$ versus $\beta$ curve is shown in Figure \ref{betatheta}.  At 30 AU, just outside the minimum focal length of 23.5 AU, the radius of the caustic is found to 
be 0.10  $\theta_E$ where $\theta_E=3.65"$, the angular Einstein ring radius. In terms of the dimensions of the Sun, the radius of the Einstein ring is 0 .10 $R_\odot$ and the caustic is 0.01 $R_\odot$. Similarly the radius of the critical curve is found to be 0.50 $\theta_E$ or $\approx$ 0.06 $ R_\odot$.  Table \ref{radii} summarizes the relative radii of Einstein ring, caustic and critical curve for different observer locations. $\theta_\odot$ is the angular measure of the Sun's radius for the corresponding $D_L$.

\clearpage

\begin{deluxetable}{lccccc}
\tabletypesize{\scriptsize}
\tablecaption{Radii of critical curves and caustics }
\tablewidth{0pt}
\tablehead{
\colhead{$D_L$(AU)} & \colhead{$\theta_E/\theta_\odot$} & \colhead{$\theta_{caus}/\theta_E $} & 
\colhead{$\theta_{crit}/\theta_E$} & \colhead{$\theta_{caus}/\theta_\odot$} & \colhead{$\theta_{crit}/\theta_\odot$} 
}
\startdata
23.0 & 0  &  0     &  0 &  0  &  0 \\
25.0    &  5.62 $10^{-2}$&  3.68  $10^{-2}$  &  0.481 &  2.07  $10^{-3}$ &  2.70 $10^{-2}$\\
27.0    &  8.24 $10^{-2}$&  6.14  $10^{-2}$&  0.518 &  5.06 $10^{-3}$&  4.28  $10^{-2}$ \\
30.0  & 11.3 $10^{-2}$& 10.0  $10^{-2}$ & 0.519 & 11.3  $10^{-3}$ & 5.87  $10^{-2}$\\
\enddata
\label{radii}
\end{deluxetable}

\clearpage

\begin{figure}[h]
\begin{center}
\includegraphics[width=.70\textwidth]
{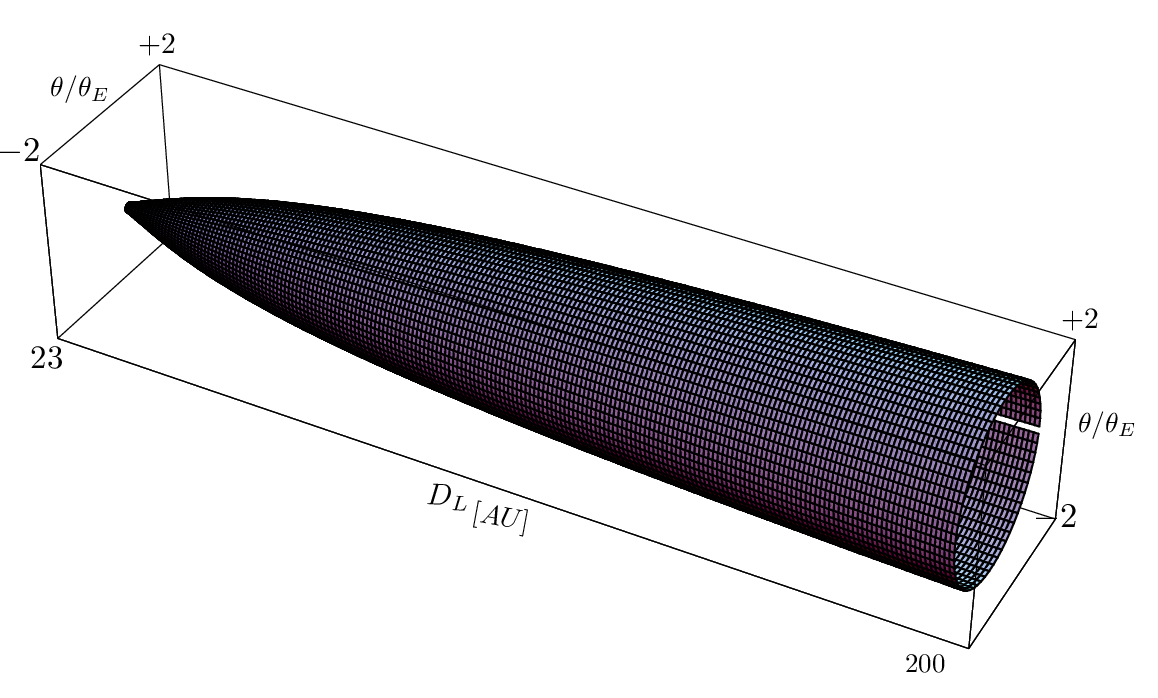}
\caption{The points inside the above surface of revolution represent the volume where an observer sees multiple images. The angular distances are normalized to the units of Einstein ring radii corresponding to a given $D_L$ along the horizontal axis.}
\label{threeimg}
\end{center}
\end{figure}

\clearpage

The apparent relative motion of an angularly small source behind the Solar lens will result in a light curve with sharp spikes when the source crosses a caustic.  The magnification at a given location can be obtained from the $\beta-\theta$ curve corresponding to that location 
\begin{equation}
\mu=\frac{\theta}{\beta}\frac{d\theta}{d\beta}.
\label{mag}
\end{equation}

\clearpage

\begin{figure}[h]
\begin{center}
\includegraphics[width=.70\textwidth]{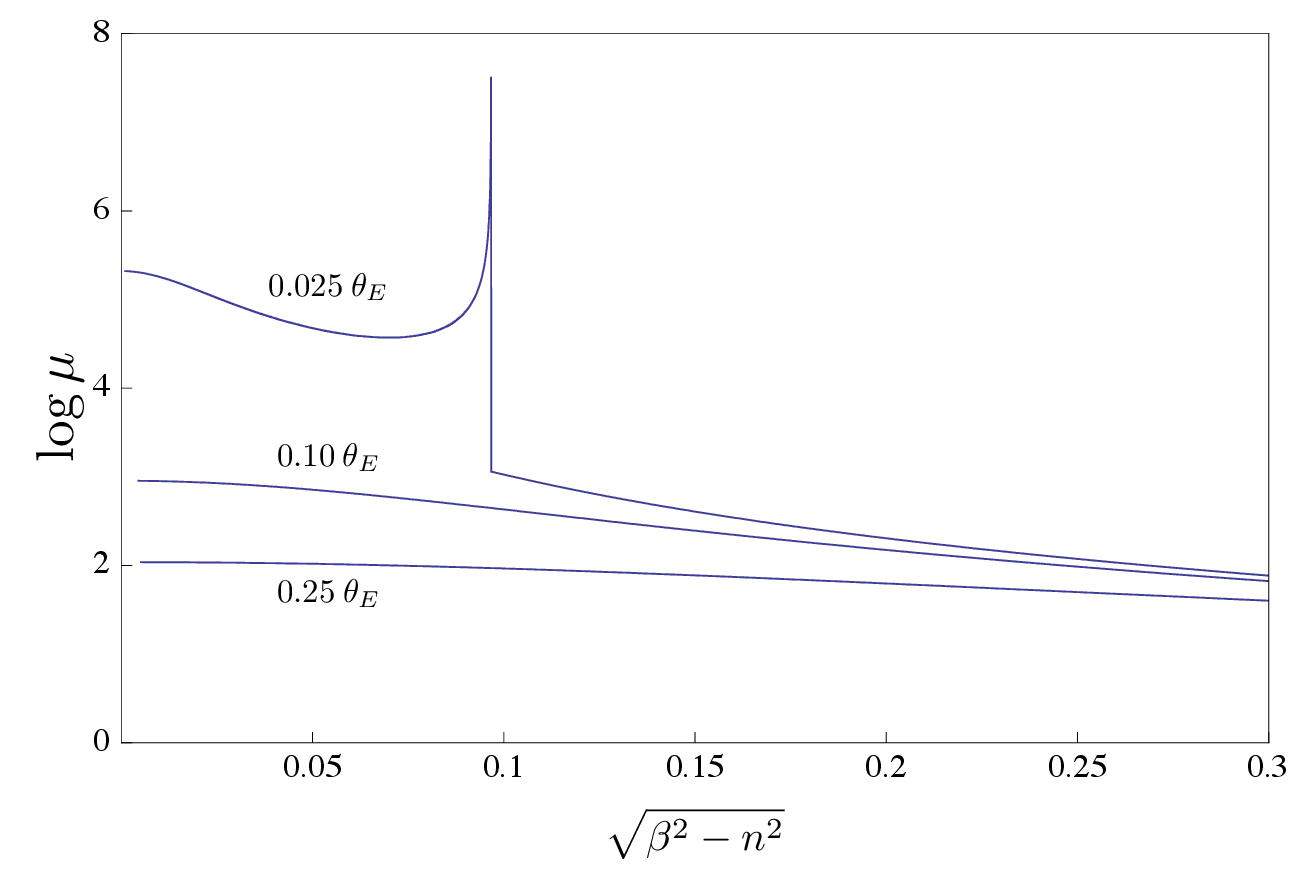}
\caption{The light curve on the top is that of a point source, moving behind the lens with an impact parameter, $n$, expressed as an angular measure of $0.025\;\theta_E$. As the distance separating the source and the lens increases, a pair of images disappears. The vertical axis is the magnification in logarithmic scale and the horizontal axis represents the distance scale normalized to angular size of the Einstein ring $\theta_E$.}
\label{figmag}
\end{center}
\end{figure}

\clearpage

The straight line motion of the source behind a weakly acting lens will result in a light curve that peaks at the minimum distance the source appears from the lens.  This minimum angular distance is 
called the impact parameter, $n$ for the source. The distance separating the projected source on the lens plane and the center of the lens scales as $\beta$ and the corresponding distance for the single image scales as $\theta$ for small $\beta$ and $\theta$.
For large distances of the source from the center of the lens, the image nearly coincides with the source. But, as this distance decreases, the image and the source locations on the lens plane diverge as seen by the observer. 
For a strong lens, unlensed source positions inside a circular caustic create three images of the source visible to the observer.  Regardless of the number of observed source images, a finite size source will undergo a finite total magnification.

The transparent Sun is a local wave zone for radiation emitted by distant sources \citep{isaacson,thorne}. Therefore, weak field limit is applicable and geometric optics can be used to calculate image locations \citep{isaacson}. 
Large magnifications allows us a more sensitive look for neutrino flux from nearby stars, as well as gravitational radiation.

The multiple images that are created in pairs when the source straddles the caustic will always lie within the Einstein radius. Therefore, they can be only separated by a distance less than or equal to the Einstein radius
\begin{equation}
R_E=\left(\frac{4GM(b)D_L}{c^2}\right)^{1/2},
\end{equation}
where $D_L$ is the distance of the observer from the lens plane. These two images (in the lens plane) must combine to produce an interference pattern with a fringe width $w\sim \lambda D_L/R_E$ at the detector \citep{nakamura}, where $\lambda$ is the wavelength of the lensed radiation. The magnification $\mu \sim R_E/w$ is appreciable only if $4GM(b)/\lambda c^2 \gg 1$. Therefore, the images cannot be resolved if the wavelength of the radiation is of the order of the Schwarzschild radius of the lens. Near the minimum focal length, diffraction effects impose even stricter constraints on $\lambda$.
However, the individual gravitational wave trajectories, magnifications and phases can be 
calculated using ray optics.


High energy neutrinos, by definition, have energies greater than 100 GeV \citep{gaisser}. The Sun admits neutrinos, unimpeded by electron scattering, only for energies up to 300 GeV \citep{escribano}. The mass distribution and chemical composition (75$\%$ hydrogen and 25$\%$ helium) of the Sun sets an upper limit on the allowed neutrino energy spectrum. Diffraction has no effect on neutrinos as the de Broglie wavelength of 300 GeV neutrino is much smaller ($\sim 10^{-15}$ m) than the gravitational radius of the lens.

\section{Discussion}

The effects of plasma and core rotation in the interior of the Sun, thought to be small, were ignored during this analysis.   The minimum focal length, the critical curve and caustics were computed numerically. The magnification of point sources for typical source separations was analyzed. The code can be modified to simulate galaxies and dark matter halos with an assumed density profile as input parameters.

To date, four spacecraft have traveled past the minimum transparent focal length of our Sun, just beyond the orbit of Uranus.  Launching a spacecraft to this distance with sufficiently sensitive gravitational wave or neutrino detectors remains a dream, however.  In a different paper, the possibility of detecting strong gravitational radiation and a wide range of neutrino energy spectrum will be examined in some detail.  With the advancement of modern technology and the improvement in the resolution power of detectors, such an endeavor could shed new light on sources emitting neutrinos, high energy gravitational waves or hitherto hypothetical particles that are predictions of at least some theories.

We thank the anonymous referee for useful suggestions, especially pointing out equation (\ref{referee}) and some of its consequences that have helped improve the paper.

\appendix
\section{Appendix}
The image locations for a given impact parameter are found by solving equation (\ref{lens}).   By allowing for a range of both positive and negative values for the deflection angle $\alpha$ for all positive values of source position $\beta$, and interpolating within bounds, one obtains the image locations $\theta$.  Therefore, this can be termed as a controlled ray tracing algorithm 
\begin{equation}
\beta=\frac{b}{D_L} ,
\end{equation}
where $b$, and $D_L$ are the impact parameter and fixed observer distance from the lens' center.  The resulting curve is shown in Figure \ref{betatheta}. For point sources the magnification can be computed directly from the curves using equation (\ref{mag}). Magnifications for extended sources can be computed 
by approximating the source as a point or inverse ray-shooting, or, occasionally using Stokes' theorem.

\begin{figure}[h]
\begin{center}
\includegraphics[width=.40\textwidth]{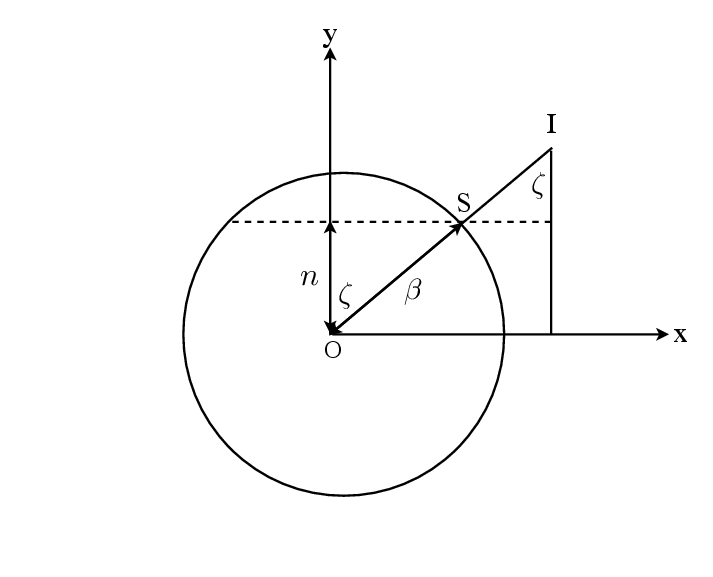}
\caption{The dashed line is the trajectory of the source ($S$) behind the lens, making an impact parameter $n$. The image ($I$) and the source always lies along a straight line for a spherical symmetric lens. Therefore, the angular distance of the source from the center of the lens along the source trajectory, as seen by the observer at any  given  time is $\sqrt{\beta^2 - n^2} $}
\label{angles}
\end{center}
\end{figure}

For a circularly symmetric lens, the deflection depends only on the impact parameter.  In that case, a new angular  measure $n$ can be defined as in Figure \ref{angles}, the perpendicular distance from the source on the lens plane.
The observer can choose a coordinate system with her $y$ coordinate and origin made to coincide with $\vec{n}$ through the center of the lens.  Now at different source positions, as seen by the observer, the source will subtend an angular measure 
\begin{equation}
x=\sqrt{\beta^2-n^2},
\end{equation}
from the $x$ axis, where $\beta\ge n$.

\clearpage

\end{document}